\documentclass[aps,prl,twocolumn,a4paper,floatfix]{revtex4} 
\usepackage{amsmath}
\usepackage{amsfonts}
\usepackage{epsfig}
\graphicspath{{figures/}}
\begin{document}
  \title{Hopping modulation in a one-dimensional Fermi-Hubbard Hamiltonian}
   \author{Francesco Massel}
   \affiliation{Department of Applied Physics, P.O. Box 5100,
                02015 Helsinki University of Technology, Finland}

   \author{Mikko J. Leskinen} 
   \affiliation{Department of Applied Physics, P.O. Box 5100,
                02015 Helsinki University of Technology, Finland}

  \author{P\"aivi T\"orm\"a}
  \email{paivi.torma@hut.fi}
   \affiliation{Department of Applied Physics, P.O. Box 5100,
                02015 Helsinki University of Technology, Finland}

  \begin{abstract}
    We consider a strongly repulsive two-component Fermi gas in a
    one-dimensional (1D) optical lattice described in terms of a
    Hubbard Hamiltonian. We analyze the response of the system to a
    periodic modulation of the hopping amplitude in presence of large
    two body interaction.
      By (essentially) exact simulations of the time evolution, we find
    a non-trivial double occupancy frequency dependence. We show how
    the dependence relates to the spectral features of the system
    given by the Bethe ansatz. The discrete nature of the spectrum is
    clearly reflected in the double occupancy after long enough
    modulation time. We also discuss the implications of the 1D
    results to experiments in higher dimensional systems.
  \end{abstract} 

  \maketitle

  Ultracold atomic gases systems couple weakly with the surrounding
  environment and are highly controllable
  \cite{jaksch__2005,chin_feshbach_2008,jaksch_cold_1998,peil_patterned_2003},
  therefore they offer excellent possibilities to investigate the
  dynamics of strongly correlated quantum many-body systems. Much
  attention has been recently devoted to the study of the dynamical
  properties both from the experimental
  \cite{jordens_mott_2008,schneider_metallic_2008} and theoretical
  \cite{kollath_modulation_2006,winkler_repulsively_2006,cramer_exact_2008,
    flesch_probing_2008, sensarma_modulation_2009} point of view.
  Especially, one-dimensional (1D) systems, accessible by experiments
  and theoretically exactly solvable in some cases, can be used to
  obtain thorough understanding of the many-body ground state and the
  dynamics. In this letter, we present an (essentially) exact
  time-evolving block decimation TEBD simulations of the dynamics of a
  repulsively interacting 1D system and reveal a non-trivial
  time-dependence which we explain using the Bethe ansatz (BA).  We
  extend the analysis also to the harmonically trapped case essential
  for ultracold gases experiments. In higher dimensions, the relevance
  of Mott and antiferromagnet (AF) physics in connection to high-$T_c$
  superconductivity (see e.g.  \cite{hur_superconductivity_2008}),
  suggests that the investigation of the equivalent systems in the
  framework of ultracold gases, especially in two dimensions, may shed
  new light on high-$T_c$ superconductor physics. Our results are
  relevant for such experiments by showing --- with analysis that does
  not assume mean-field approximation nor linear response --- how the
  discrete nature of the spectrum is reflected in the dynamics of a
  lattice modulation experiment.

  We examine the dynamical properties of a two-species ultracold
  atomic gas loaded in a 1D optical lattice both for open boundary
  conditions (obc) and in presence of parabolic confinement.  In
  particular we perform an (essentially) exact numerical simulation of
  this system when a periodic lattice modulation is applied.  To this
  end we consider the 1D Hubbard Hamiltonian, in presence of an
  external parabolic confining potential
   \begin{equation}
     \label{eq:HubbHam}
    H=H_J+H_{int}+\sum_i^L V_i (n_{i\,\uparrow}+n_{i\,\downarrow}),
  \end{equation}
  where $H_J=-J\sum_{i,\sigma=\{\uparrow,\downarrow\}}^L
  c^\dagger_{i\,\sigma} c_{i+1\,\sigma} + h.c. $, $H_{int}=U\sum_i^L
  n_{i\,\uparrow}n_{i\,\downarrow}$, $V_i=\Omega(i-i_0)^2$, $J$ is the
  hopping amplitude, and $U$ the on-site interaction. In the limit
  $U/J\gg 1$, the only effect of the lattice modulation is a
  modulation of the hopping amplitude $J$, the effect on $U$ being
  negligible (see \cite{sensarma_modulation_2009}).
    
  We focus on the double occupancy (d.o.) expectation value
  $<n_{i\,\uparrow}n_{i\,\downarrow}>$, when a (small) periodic
  modulation $ \delta J \sin (\omega t)$ of the hopping amplitude $J$
  is applied for a given time to the ground state of the Hamiltonian
  given in Eq.  \eqref{eq:HubbHam} for two different situations:
  $U/J=60$, $L=20$, particle number $N_p=12$ and $U/J=20$, $L=40$,
  $N_p=24$ respectively.  Our simulation is performed with a TEBD
  algorithm \cite{vidal_efficient_2003,vidal_efficient_2004}, both for
  the ground-state calculation (imaginary-time evolution) and the
  real-time evolution.  The numerical results show that the ground
  state is constituted by a central Mott region with one atom per
  site, surrounded by two small metallic (Luttinger liquid) regions
  where the filling is less than one. In order to avoid finite-size
  effects, we have considered a lattice size exceeding the actual
  extent of the atomic cloud by a few lattice sites. Heuristically,
  the Luttinger liquid phase corresponds to the regions where
  $<n_i^2>-<n_i>^2\neq const$, see \cite{rigol_local_2003}. Moreover,
  the static structure factor $S(q)$ in the ground state of the finite
  systems here considered exhibits the same qualitative features of
  the 1D Heisenberg AF chain (i.e. a slow decaying peak centered
  around $q=\pi$), as expected in the $U\gg J$ limit of the Hubbard
  Hamiltonian \cite{essler_one-dimensional_2005}.
    
  For an infinite chain at half filling, if hopping and parabolic
  confinement are suppressed, the (highly degenerate) first excited
  state is represented by a site with an empty site and a doubly
  occupied one (henceforth particle-hole excitation).  The energy gap
  between this state and the ground state is equal to $U$. To
  investigate the spectral properties of the system when $U \gg J$, it
  seems then natural to choose $\omega \simeq U$ in $ \delta J \sin
  (\omega t)$, with $\delta J/J=0.1$ throughout the paper.
  \begin{figure}
    \centering
    \epsfig{file=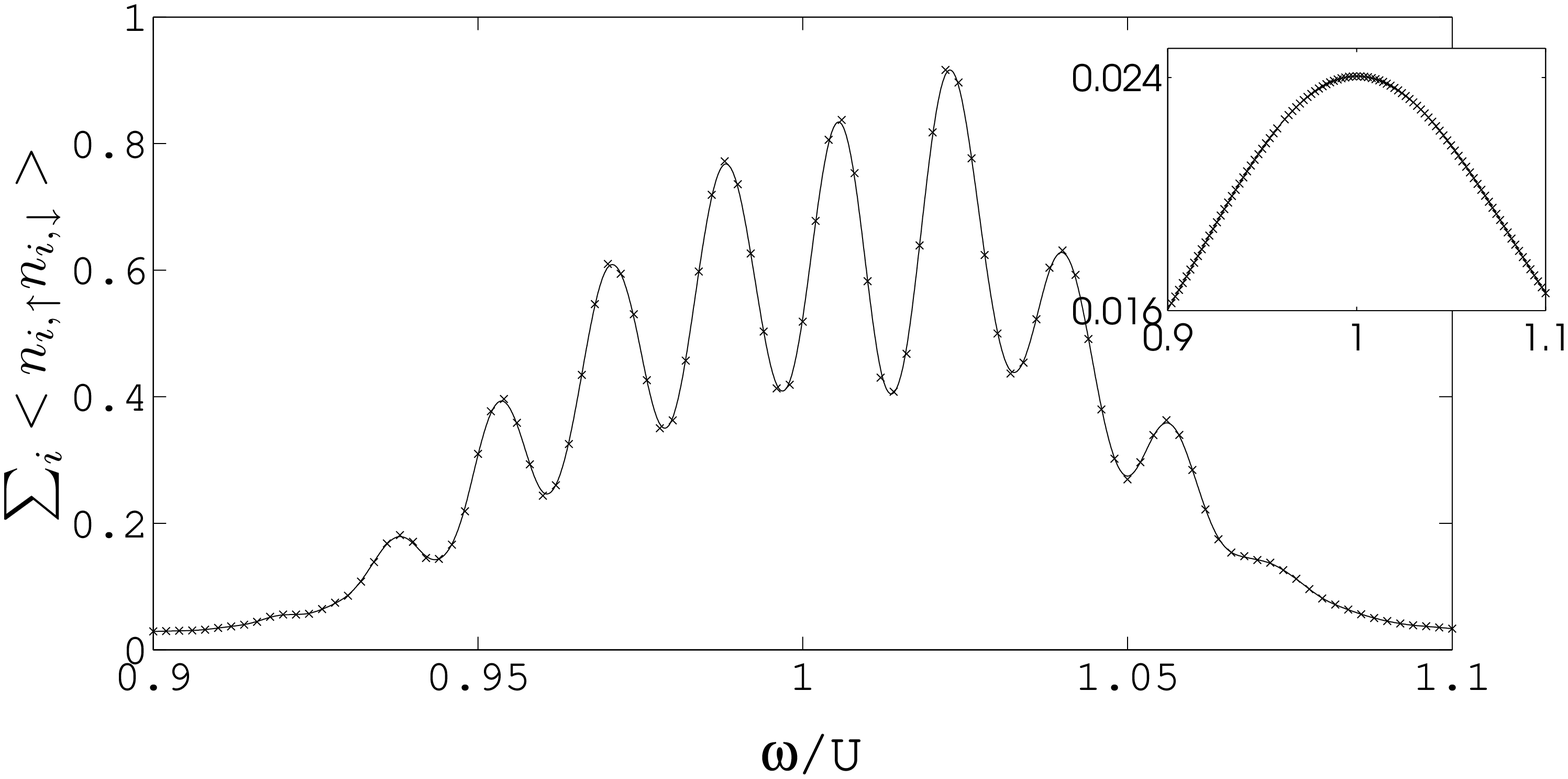,width=0.5\columnwidth}
    \caption{Double occupancy $\sum_i<n_{i,\uparrow}n_{i,\downarrow}>$
      as a function of frequency for long ($t=10$ main panel) and short
      ($t=0.5$ inset) times. A single broad peak appears for $\omega/U \simeq
      1$ ($L=20$, $U/J=60$, $\Omega/J=0.1$).}
   \label{fig:do}
  \end{figure}
  \begin{figure}
    \centering
    \epsfig{file=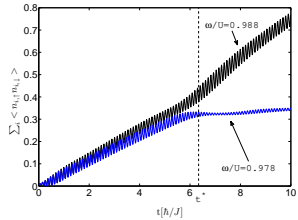,width=0.5\columnwidth}
    \caption{(color online) Double occupancy
      $\sum_i<n_{i,\uparrow}n_{i,\downarrow}>$ as a function of time, for
      two different frequency values corresponding to a minimum
      ($\omega/U=0.978$) and a maximum ($\omega/U=0.988$) of the plot in
      Fig.  \ref{fig:do} (same parameters as Fig. \ref{fig:do}).}
   \label{fig:do_t}
  \end{figure}

  In Fig. \ref{fig:do} (inset) we show the d.o. as a
  function of frequency for short times, from which one notices that a
  broad peak appears as in correspondence of the value $\Omega = U$,
  consistently with the value of the gap of the particle-hole
  excitation. However, for larger times a richer structure appears
  (Fig.  \ref{fig:do}). From Fig.  \ref{fig:do_t}, it is clear how it
  is possible to distinguish between short- ($t < t^*$) and long- ($t
  > t^*$) time behavior, the explicit value of the threshold $t^*$
  will be derived below.

  The key idea in understanding the time dependence of
  $<n_{i\,\uparrow}n_{i\,\downarrow}>$ is that the resolution of
  degeneracy depends on the modulation timescale. For $t<t^*$ it is
  possible to consider the particle-hole excited states as
  quasi-degenerate, hence the transition probability between the
  ground state and the quasi-continuum (centered around $U$) of
  excited states is given by $
    P(t)\propto \delta J^2 
     \sin^2\left[\left(U-\hbar\omega \right)t/\,2\hbar \right]\
     (U-\hbar\omega)^2. $
  The threshold time is then calculated as the time when the internal
  structure of the excited states band becomes visible, namely $ t^* =
  2 \pi/\Delta E$, where $\Delta E$ is the energy difference in the
  quasi continuum of excited states which, in our case, will be
  determined by BA. For $t>t^*$, the transition probability
  is given by
  \begin{equation}
    \label{eq:sumc}
    \sum_{n,\,E_{n}\simeq E_{gs}\pm \hbar \omega}P_n(t)=
    \left(\frac{2\pi}{\hbar}\right)|V_{n,\,gs}|^2
      \rho(E_{n})t\Bigg|_{{E_n}\simeq E_{gs}\pm \hbar \omega},
  \end{equation}
  where $E_n$ represents the energy of one of the quasi-degenerate
  excited states. The numerical results plotted in Fig. \ref{fig:do_t}
  show how the transition between these two regimes is quite abrupt,
  allowing an easy comparison with the theoretical value of $t^*$.
  Eq. \eqref{eq:sumc} explains the piecewise linear time dependence of
  d.o., which is induced by the different effective density of states
  $\rho (E_n)$ and perturbation matrix elements $|V_{n,gs}(\delta
  J)|^2$ before and after $t^*$. In particular, in order to have a
  peak in the d.o. spectrum, both the perturbation matrix element and
  the density of states at the appropriate energy must be different
  from zero, suggesting that the d.o. spectrum exhibits selection
  rules in its long-time peak distribution. In the 2-site problem, it
  is possible to see explicitly how the ground state with two
  particles, represented by singlet state $|S>
  =\frac{1}{\sqrt{2}}\left( | \uparrow, \downarrow> - | \downarrow,
    \uparrow> \right)$, is coupled by the hopping modulation to the
  state $|D_+>= \frac{1}{\sqrt{2}}\left( | \uparrow \downarrow,0> + |
    0, \uparrow \downarrow> \right)$ only, and not to the state
  $|D_->=\frac{1}{\sqrt{2}}\left( | \uparrow \downarrow,0> - | 0,
    \uparrow \downarrow> \right)$.

  The 1D nature of the problem allows some insight on the peak
  position deriving from the exact (BA) solution of the 1D Hubbard
  Hamiltonian. As a first step we make contact between BA and the
  numerical solution of a small linear chain with open boundary
  conditions ($\Omega=0$) and $U/J \to \infty\,(U/J=500)$.  To this
  end we consider the BA equations for an open Hubbard chain (see
  \cite{asakawa_finite-size_1996} and EPAPS supplementary material
  \cite{EPAPS}).   
  The solution of BA equations with
  respect to the charge momenta $k_j$ and spin rapidities
  $\lambda_\alpha$ allows to determine energy and momentum eigenstates
  whose values can be expressed in terms of $k_j$ as 
  \begin{equation}
    \label{eq:spec}
     E=-2J\sum_{j=1}^N \cos(k_j) \qquad P=\left[ \sum_{j=1}^N k_j \right]
  mod\, 2\pi.   
  \end{equation} 
  Along the lines of the derivation by Ogata and Shiba
  \cite{ogata_bethe-ansatz_1990} applied to the case of OBC,
  Bethe-ansatz equations lead to the simple relation
  $k_j=\pi\,I_j/(L+1)$, in the limit $U/J\to \infty$. The latter
  expression is particularly relevant since it describes the spectrum
  of the considered Hamiltonian in terms of spinless fermionic
  particles.  This result can be interpreted within the general
  framework of spin-charge separation of excitations in 1D systems
  (see e.g.  \cite{giamarchi_quantum_2003}). In particular the ground
  state for $N=L$ is obtained when $I_j=1 \dots N$. Its energy is
  given by $ E=-2J\sum_{j=1}^L \cos(k_j) $ and the Fermi
  quasi-momentum is given by $ k_F=\pi\,L/(L+1)$.  The excitations of
  the system can be described in terms of particle-hole excitations of
  this system.  Hence, since the first Hubbard band is full,
  \begin{equation}
    \label{eq:Exc}
     \tilde{\Delta} E=\Delta E +U=-2J (\cos(k_p)-\cos(k_h))+U,
  \end{equation}
  where $E_p=-2J \cos k_p +U$ corresponds to the energy of an extra
  particle added in the second Hubbard band, and $2J\cos(k_h)$ the
  energy of the hole in the first Hubbard band. Since having a
  particle and a hole in the same momentum state would not contribute
  to the increase of the interaction energy by $U$, these states must
  not be considered in the calculation of the energy level structure
  around $U$. We also note that in the thermodynamic limit
  $L\rightarrow\infty$, the discrete energy-level structure becomes a
  continuous band of width $8J/U$, as it is possible to deduce from
  Eq.\eqref{eq:Exc}. Our simulations are at $T=0$, but the features in
  the spectra are expected to be visible if the system is in the Mott
  state and $T < \tilde{\Delta} E$.

\begin{figure}
    \centering
    \epsfig{file=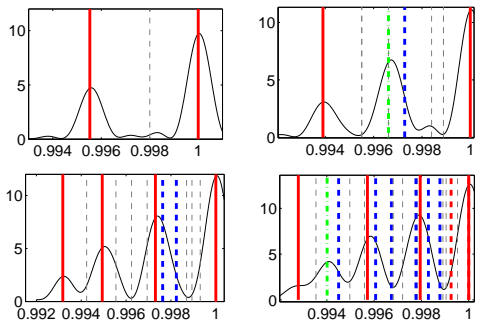,width=0.9\columnwidth}
    \caption{(color online) d.o. spectrum for $U/J=500$ and
      $L=4,6,8,10$ (top-left to bottom right). Vertical lines
      represent the value $\tilde{\Delta} E$ for various $k_p$ and
      $k_h$. Red continuous lines represent the values listed in Table
      \ref{tab:ph}.  Blue dashed lines represent transitions between
      states of even parity (i.e.  $i_p$ and $i_h$ even), which do not
      seem to correspond to a peak in the d.o.  spectrum. Grey dotted
      lines represent states with $i_p=2n+1$ or $i_h=2n+1$, with $n
      \in \mathbb{Z}$ . Green dash dotted lines represent a particular
      condition of coincidence of the frequency between a ``red line''
      transition and a ``gray line'' transition with $i_p=2n+1$,
      $i_h=2n'+1$ and $i_p-i_h=2n''$ (see Table \ref{tab:ph}). For ease
      of reading, the right part of the spectrum has been omitted due
      to its symmetry around $\omega=1$ (c.f. the slight asymmetry in
      the trapped case, Fig.  \ref{fig:do} ). x axis: $\omega/U$, y axis:
      $<n_{i\,\uparrow}n_{i\,\downarrow}>$).}
   \label{fig:nud_sp}
  \end{figure}

  We have made a connection to the particle-hole excitation spectrum
  by performing hopping modulation on systems with open boundary
  conditions (i.e. no parabolic confinement), with $U/J=500$ and
  $L=4,\,6,\,8,\,10$. As seen in Fig.  \ref{fig:nud_sp}, the d.o.
  peaks correspond to specific particle-hole excitations, supporting
  the explanation of the d.o. spectrum in terms of selection rules
  (see Table \ref{tab:ph}).
  \begin{table} \scriptsize
    \caption{Theory/numerics comparison}
    \centering
      \begin{tabular}{c c c c c}
     & $i_p$&$i_h$&$ \tilde{\Delta} E/U$&$(\tilde{\Delta} E-\omega_{pk})/U$\\
      \hline
    ($L=4$)&  4    &  4  &  1.0000  &   $<$1e-4             \\
           &  2    &  4  &  0.9955  &   $<$1e-4             \\ 
           & (1    &  3  &  0.9955  &   $<$1e-4 )$\,^*$     \\ 
      \hline 
    ($L=6$)&  6    &  6  & 1.0000   &   $<$1e-4              \\
           &  2    &  4  & 0.9966   &   $\simeq$ 1e-4        \\ 
           & (3    &  5  & 0.9966   &   $\simeq$ 1e-4 )$\,^*$\\ 
           &  2    &  6  & 0.9939   &   $<$1e-4              \\              
      \hline 
    ($L=8$)& 8    &  8  & 1.0000   &   $<$1e-4               \\
           & 4    &  6  & 0.9973   &   $\simeq$ 1e-4         \\ 
           & 2    &  6  & 0.9949   &   $<$1e-4               \\ 
           &(3    &  7  & 0.9949   &   $<$1e-4)$\,^*$        \\  
           & 2    &  8  & 0.9931   &   $<$1e-4               \\                 
      \hline
   ($L=10$)& 10   &  10 & 1.0000   &   $<$1e-4               \\
           & 6    &  8  & 0.9979   &   $<$1e-4               \\
           & 4    &  8  & 0.9957   &   $\simeq$ 2e-4         \\ 
           & 2    &  8  & 0.9940   &   $<$1e-4               \\
           &(3    &  9  & 0.9940   &   $<$1e-4)$\,^*$        \\
           & 2    &  10 & 0.9928   &   $<$1e-4               \\ 
      \hline
    \end{tabular}
   \label{tab:ph}
  \end{table}
  Even if the simulations in presence of a parabolic confining
  potential with $U/J=20,\,60$ cannot be described exactly within the
  above formalism, we suggest that it is possible to make contact
  between a description in terms of spinless fermions and the
  numerical results, the only effect of parabolic confinement being
  the different spectrum in the Mott phase.  This hypothesis is
  justified by the nature of the metallic phase in the limit $U/J\gg
  1$ (gapless spectrum, small spatial extent) and confirmed by
  numerical evidence.  As in the case where the parabolic confining
  potential is absent, we assume that the spectrum of the system can
  be described in terms of particle-hole excitations of spinless
  fermions moving in a lattice, in presence of a global parabolic
  confinement.

  In \cite{rey_ultracold_2005}, it has been shown that, for $4J
  \gtrsim \Omega$, the approximate description of the single-particle
  spectrum can be carried out in terms of low-energy excitations with
  quantum number $n<n_c$, $n_c=2\|\sqrt{2J \Omega}\|-1$, where $\|x\|$
  is the integer closest to $\|x \|$, and high-energy excitations with
  $n>n_c$.  If $n<n_c$, the single particle energy is given by
  \begin{multline}
    \label{eq:En_ho}
    E_n-E_0=2 \sqrt{J \Omega } (n+1/2)- \\ \frac{\Omega}{32} 
              \left[(2n+1)^2+1-\frac{(2n+1)^3+3(2n+1)}{32\sqrt{J/\Omega}}\right],
  \end{multline}
  where $E_0$ is a constant energy term.  If, on the other hand, $n
  \geq n_c$, for the high energy modes the energy eigenvalues are
  given by
  \begin{equation}
    \label{eq:En_loc}
    E_{n=2i} \simeq E_{n=2i-1} \simeq \Omega i^2 + \frac{2J}{(2i)^2-1}, 
  \end{equation}
  representing states close to position eigenstates.   
   
\begin{figure}
    \centering
    \epsfig{file=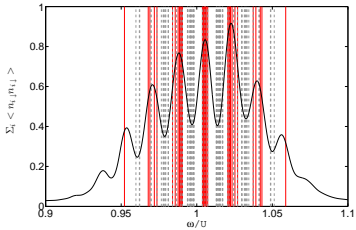,width=0.6\columnwidth}
    \caption{(color online) Trapped system: energy levels (continuous
      red lines correspond to even values of $i_p$, $i_h$) for $L=20$,
      $N_{tot}=12$, $U/J=60$, $\Omega/J=0.1$. The peaks at the edge of
      the diagram do not correspond to any peak due to the approximate
      description of the energy levels.}
   \label{fig:EnResPar}
  \end{figure}

  In the trapped system, performing a calculation analogous to the case
  with open boundary conditions, it is now possible to compare the
  particle-hole excitation spectrum to the d.o.  spectrum obtained
  numerically (see Fig. \ref{fig:EnResPar}). It must be noted that, in
  Fig.  \ref{fig:EnResPar}, we have shifted ($\Delta\omega/U \simeq
  0.667$) the energy spectrum calculated in Eqs.  \eqref{eq:En_ho},
  \eqref{eq:En_loc}, in order to obtain the superposition with the
  d.o.  spectrum. The shift is due to finite value of $U$. This effect
  can be explicitly calculated in the two-site case, where the
  energy-level splitting can be obtained analytically. In the case
  $U/J=60$, the d.o. spectrum shows good agreement with the spectrum
  from Eqs. \eqref{eq:En_ho} and \eqref{eq:En_loc}, when we enforce
  the condition $n_p=even$, $n_h=even$.
  \begin{figure}
    \centering
    \epsfig{file=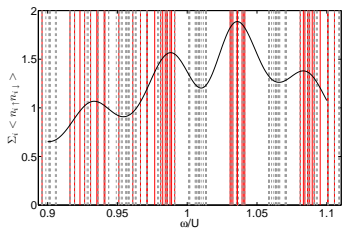,width=0.6\columnwidth}
    \caption{(color online) Trapped system: energy levels (same
      definitions as in Fig.  \ref{fig:EnResPar}) $L=40$, $N=24$,
      $U=20$, $\Omega/J=0.1$ (right), when values of $i_h$ and $i_p$ up
      to 20 are considered.}
   \label{fig:EnResPar2}
  \end{figure}

  Our analysis of the dynamical properties of a 1D Hubbard chain is in
  close relation to the experimental \cite{jordens_mott_2008} and
  theoretical \cite{sensarma_modulation_2009} investigations of Mott
  and AF phases in optical lattices.  Even if in
  our simulations there is no proper AF phase, due to the
  dimensionality of the system, it is possible to relate the results
  obtained here concerning the peaks in the d.o. spectrum to those
  appearing in \cite{sensarma_modulation_2009}. In presence of AF
  ordering, an energy gap $\propto J^2/U$ develops. In our case, the
  separation between the different energy levels, which mimics the
  effects of an energy gap, is induced either by the parabolic
  confinement or by finite size effects. The qualitative similarity
  between the spectrum obtained here and in
  \cite{sensarma_modulation_2009} confirms the relationship between
  the presence of finite energy gaps and the peaks in the d.o.
  spectrum. In our case, especially for the obc case and for $U=60$,
  we are able to suggest how the excitation spectrum together with the
  nature of the lattice modulation determines the structure of the
  peaks.  While \cite{sensarma_modulation_2009} considers a higher
  dimensional system thus closer to the present experiments, our study
  is complementary in the sense that it does not assume mean-field
  approximation or linear response.

  In addition, when performing a d.o.  modulation experiment aiming at
  the observation of the spectrum depicted in Fig. \ref{fig:do}, the
  effect of a finite modulation time must be taken into account. If we
  consider experimentally relevant values (see
  \cite{jordens_mott_2008}) of $U$, $J$ and the modulation time
  $t_{mod}$ ($U \simeq 5 \text{kHz}$, $J\simeq 0.1 \text{kHz}$,
  $t_{mod}=50/U=0.01\,\text{s}$ ), we obtain $t^* \simeq 2\pi U/J^2
  \simeq 3.14\,s$. Since this is experimentally rather long time, the
  estimate suggests that smaller values of $U/J^2$ might be necessary
  to observe the AF gap. In this calculation of $t^*$ we have
  considered $\Delta_{AF} \simeq J^2/U$.

  In summary, with the aid of a TEBD numerical simulation, we have
  analyzed the properties of a fermionic gas in a 1D optical lattice,
  in presence of parabolic confinement. The numerical results show
  that, for a sufficiently long time, a nontrivial peak structure
  appears in the d.o. spectrum. The peak structure has been
  qualitatively justified in terms of the BA solution of a chain with
  open boundary conditions, in the limit of $U/J \to \infty$.  The
  results we have obtained, while not being a direct evidence of what
  should be observed in a modulation experiment in 3D
  \cite{jordens_mott_2008} when $T < T_{\textrm{N\'eel}}$
  \cite{neel_1936} , due to the different nature of the gaps in 3D and
  our 1D systems, explain in detail how the discrete nature of the
  spectrum is reflected in the lattice modulation experiment.  The
  results thus suggest that the gap in the AF phase can be observed
  via a nontrivial peak structure.

  This work was supported by Academy of Finland and EuroQUAM/FerMix
  (Projects No. 213362, 217041, 217043, and 210953) and conducted as
  part of a EURYI scheme grant \cite{esf}.


\begin{thebibliography}{22}
\expandafter\ifx\csname natexlab\endcsname\relax\def\natexlab#1{#1}\fi
\expandafter\ifx\csname bibnamefont\endcsname\relax
  \def\bibnamefont#1{#1}\fi
\expandafter\ifx\csname bibfnamefont\endcsname\relax
  \def\bibfnamefont#1{#1}\fi
\expandafter\ifx\csname citenamefont\endcsname\relax
  \def\citenamefont#1{#1}\fi
\expandafter\ifx\csname url\endcsname\relax
  \def\url#1{\texttt{#1}}\fi
\expandafter\ifx\csname urlprefix\endcsname\relax\def\urlprefix{URL }\fi
\providecommand{\bibinfo}[2]{#2}
\providecommand{\eprint}[2][]{\url{#2}}

\bibitem[{\citenamefont{Jaksch and Zoller}(2005)}]{jaksch__2005}
\bibinfo{author}{\bibfnamefont{D.}~\bibnamefont{Jaksch}} \bibnamefont{and}
  \bibinfo{author}{\bibfnamefont{P.}~\bibnamefont{Zoller}},
  \bibinfo{journal}{Ann. of Phys.} \textbf{\bibinfo{volume}{315}},
  \bibinfo{pages}{52} (\bibinfo{year}{2005}).

\bibitem[{\citenamefont{Chin et~al.}(2008)}]{chin_feshbach_2008}
\bibinfo{author}{\bibfnamefont{C.}~\bibnamefont{Chin}} \bibnamefont{et~al.},
  \bibinfo{journal}{arXiv:0812.1496}  (\bibinfo{year}{2008}).

\bibitem[{\citenamefont{Jaksch et~al.}(1998)}]{jaksch_cold_1998}
\bibinfo{author}{\bibfnamefont{D.}~\bibnamefont{Jaksch}} \bibnamefont{et~al.},
  \bibinfo{journal}{Phys. \ Rev. \ Lett.} \textbf{\bibinfo{volume}{81}},
  \bibinfo{pages}{3108} (\bibinfo{year}{1998}).

\bibitem[{\citenamefont{Peil et~al.}(2003)}]{peil_patterned_2003}
\bibinfo{author}{\bibfnamefont{S.}~\bibnamefont{Peil}} \bibnamefont{et~al.},
  \bibinfo{journal}{Phys. \ Rev. A} \textbf{\bibinfo{volume}{67}},
  \bibinfo{pages}{051603(R)} (\bibinfo{year}{2003}).

\bibitem[{\citenamefont{Jordens et~al.}(2008)}]{jordens_mott_2008}
\bibinfo{author}{\bibfnamefont{R.}~\bibnamefont{Jordens}} \bibnamefont{et~al.},
  \bibinfo{journal}{Nature} \textbf{\bibinfo{volume}{455}},
  \bibinfo{pages}{204} (\bibinfo{year}{2008}).

\bibitem[{\citenamefont{Schneider et~al.}(2008)}]{schneider_metallic_2008}
\bibinfo{author}{\bibfnamefont{U.}~\bibnamefont{Schneider}}
  \bibnamefont{et~al.}, \bibinfo{journal}{Science}
  \textbf{\bibinfo{volume}{322}}, \bibinfo{pages}{1520} (\bibinfo{year}{2008}).

\bibitem[{\citenamefont{Kollath et~al.}(2006)}]{kollath_modulation_2006}
\bibinfo{author}{\bibfnamefont{C.}~\bibnamefont{Kollath}} \bibnamefont{et~al.},
  \bibinfo{journal}{Phys. \ Rev. A} \textbf{\bibinfo{volume}{74}},
  \bibinfo{pages}{041604(R)} (\bibinfo{year}{2006}).

\bibitem[{\citenamefont{Winkler et~al.}(2006)}]{winkler_repulsively_2006}
\bibinfo{author}{\bibfnamefont{K.}~\bibnamefont{Winkler}} \bibnamefont{et~al.},
  \bibinfo{journal}{Nature} \textbf{\bibinfo{volume}{441}},
  \bibinfo{pages}{853} (\bibinfo{year}{2006}).

\bibitem[{\citenamefont{Cramer et~al.}(2008)}]{cramer_exact_2008}
\bibinfo{author}{\bibfnamefont{M.}~\bibnamefont{Cramer}} \bibnamefont{et~al.},
  \bibinfo{journal}{Phys. \ Rev. \ Lett.} \textbf{\bibinfo{volume}{100}},
  \bibinfo{pages}{030602} (\bibinfo{year}{2008}).

\bibitem[{\citenamefont{Flesch et~al.}(2008)}]{flesch_probing_2008}
\bibinfo{author}{\bibfnamefont{A.}~\bibnamefont{Flesch}} \bibnamefont{et~al.},
  \bibinfo{journal}{Phys. \ Rev. A} \textbf{\bibinfo{volume}{78}},
  \bibinfo{pages}{033608} (\bibinfo{year}{2008}).

\bibitem[{\citenamefont{Sensarma et~al.}(2009)}]{sensarma_modulation_2009}
\bibinfo{author}{\bibfnamefont{R.}~\bibnamefont{Sensarma}}
  \bibnamefont{et~al.}, \bibinfo{journal}{0902.2586}  (\bibinfo{year}{2009}).

\bibitem[{\citenamefont{Hur and Rice}(2008)}]{hur_superconductivity_2008}
\bibinfo{author}{\bibfnamefont{K.~L.} \bibnamefont{Hur}} \bibnamefont{and}
  \bibinfo{author}{\bibfnamefont{T.~M.} \bibnamefont{Rice}},
  \bibinfo{journal}{Ann. Phys. (N.Y.)} \textbf{\bibinfo{volume}{324}},
  \bibinfo{pages}{1452} (\bibinfo{year}{2009}).

\bibitem[{\citenamefont{Vidal}(2003)}]{vidal_efficient_2003}
\bibinfo{author}{\bibfnamefont{G.}~\bibnamefont{Vidal}},
  \bibinfo{journal}{Phys. \ Rev. \ Lett.} \textbf{\bibinfo{volume}{91}},
  \bibinfo{pages}{147902} (\bibinfo{year}{2003}).

\bibitem[{\citenamefont{Vidal}(2004)}]{vidal_efficient_2004}
\bibinfo{author}{\bibfnamefont{G.}~\bibnamefont{Vidal}},
  \bibinfo{journal}{Phys. \ Rev. \ Lett.} \textbf{\bibinfo{volume}{93}},
  \bibinfo{pages}{040502} (\bibinfo{year}{2004}).

\bibitem[{\citenamefont{Rigol et~al.}(2003)}]{rigol_local_2003}
\bibinfo{author}{\bibfnamefont{M.}~\bibnamefont{Rigol}} \bibnamefont{et~al.},
  \bibinfo{journal}{Phys. \ Rev. \ Lett.} \textbf{\bibinfo{volume}{91}},
  \bibinfo{pages}{130403} (\bibinfo{year}{2003}).

\bibitem[{\citenamefont{Essler et~al.}(2005)}]{essler_one-dimensional_2005}
\bibinfo{author}{\bibfnamefont{F.~H.~L.} \bibnamefont{Essler}}
  \bibnamefont{et~al.}, \emph{\bibinfo{title}{The One-Dimensional Hubbard
  Model}} (\bibinfo{publisher}{Cambridge University Press},
  \bibinfo{year}{2005}).

\bibitem[{\citenamefont{Asakawa and Suzuki}(1996)}]{asakawa_finite-size_1996}
\bibinfo{author}{\bibfnamefont{H.}~\bibnamefont{Asakawa}} \bibnamefont{and}
  \bibinfo{author}{\bibfnamefont{M.}~\bibnamefont{Suzuki}},
  \bibinfo{journal}{JPA} \textbf{\bibinfo{volume}{29}}, \bibinfo{pages}{225}
  (\bibinfo{year}{1996}).

\bibitem[{\citenamefont{Ogata and Shiba}(1990)}]{ogata_bethe-ansatz_1990}
\bibinfo{author}{\bibfnamefont{M.}~\bibnamefont{Ogata}} \bibnamefont{and}
  \bibinfo{author}{\bibfnamefont{H.}~\bibnamefont{Shiba}},
  \bibinfo{journal}{Phys. \ Rev. B} \textbf{\bibinfo{volume}{41}},
  \bibinfo{pages}{2326} (\bibinfo{year}{1990}).

\bibitem[{\citenamefont{Giamarchi}(2003)}]{giamarchi_quantum_2003}
\bibinfo{author}{\bibfnamefont{T.}~\bibnamefont{Giamarchi}},
  \emph{\bibinfo{title}{Quantum Physics in One Dimension}}
  (\bibinfo{publisher}{Oxford University Press}, \bibinfo{year}{2003}).

\bibitem[{\citenamefont{Rey et~al.}(2005)}]{rey_ultracold_2005}
\bibinfo{author}{\bibfnamefont{A.~M.} \bibnamefont{Rey}} \bibnamefont{et~al.},
  \bibinfo{journal}{Phys. \ Rev. A} \textbf{\bibinfo{volume}{72}},
  \bibinfo{pages}{033616} (\bibinfo{year}{2005}).

\bibitem[{\citenamefont{Neel}(1936)}]{neel_1936}
   \bibinfo{author}{\bibfnamefont{J.} \bibnamefont{N\'eel}},
   \bibinfo{journal}{Ann. Phys. (Paris)} \textbf{\bibinfo{volume}{5}},
   \bibinfo{pages}{232} (\bibinfo{year}{1936}).

 \bibitem[22]{EPAPS} \bibinfo{url} See EPAPS Document No.
   E-PRLTAO-103-077934 for supplementary material on Bethe-ansatz
   equations for open-boundary conditions. For more information on
   EPAPS, see http://www.aip.org/pubservs/epaps.html.

\bibitem[23]{esf}
    \bibinfo{url} www.esf.org/euryi. 

 \end{thebibliography}
\end{document}